# NUMERICAL STUDIES ON LOCALLY DAMPED STRUCTURES[*]

Z. Li, N.T. Folwell and T.O. Raubenheimer, SLAC, Stanford University, Stanford, CA 94039, USA

*Abstract*

In the JLC/NLC X-band linear collider, it is essential to reduce the long-range dipole wakefields in the accelerator structure to prevent beam break up (BBU) and emittance degradation. The two methods of reducing the long-range wakefields are detuning and damping. Detuning reduces the wakefields rapidly as the dipole modes de-cohere but, with a finite number of modes, the wakefield will grow again as the modes re-cohere. In contrast, damping suppresses the wakefields at a longer distance. There are two principal damping schemes: synchronous damping using HOM manifolds such as that used in the RDDS1 structure and local damping similar to that used in the CLIC structure. In a locally damped scheme, one can obtain almost any Q value, however, the damping can have significant effects on the accelerating mode. In this paper, we present a medium local-damping scheme where the wakefields are controlled to meet the BBU requirement while minimizing the degradations of the fundamental rf parameters. We will address the load design and pulse heating issues associated with the medium damping scheme.

## 1 INTRODUCTION

The Next Linear Collider (JLC/NLC) [1,2] will provide a luminosity of $10^{34} cm^{-2} sec^{-1}$ or higher at a center of mass of 1 TeV. To obtain this luminosity, the NLC linacs must accelerate bunch trains with lengths of 266 ns and currents that are over half an ampere while preserving the very small beam emittances. Dipole wakefields in the accelerator structures can cause beam emittance degradation and beam-break-up. It is essential in the collider design to reduce the long-range dipole wakefields to prevent the beam breakup instability (BBU). The two methods of reducing the long-range wakefields are detuning and damping of the dipole modes. Detuning can reduce the wakefields rapidly by causing the modes to de-cohere however it is less effective at longer time scales because the finite number of modes can re-cohere. In contrast, damping can be used to control the wakefields at a longer distance.

There are two primary damping schemes for traveling wave structures: the synchronous coupling scheme as used in the RDDS1[3] design and the local damping scheme used in the CLIC[4] structure design. In the local damping scheme, each cell is coupled to four radial wave-guides via coupling irises. Heavy damping of the dipole mode can be achieved with the proper design of the iris openings however this can have significant effects on the fundament mode and cause reductions of $Q_0$ and the shunt impedance. As was demonstrated in the CLIC 15 GHz test structure [4] where the dipole $Q_1$ was about 20, the $Q_0$ reduction of the fundamental mode, due to the damping irises, can be as high as 20%. This reduces the rf efficiency and may lead to unwanted effects such as pulsed heating.

For the NLC X-band linacs, we propose a medium damping scheme with a $Q_1$ of a few hundred as means of solving the rf efficiency and the pulse heating limitations. In this paper, we address some of the design issues for locally damped structures at X-band. We will also discuss the design of compact loads for the medium damped structures to minimize the physical size associated with the local damping and thereby reduce the cost.

## 2 LOCALLY DAMPED STRUCTURE

The numerical studies in this paper will be based on a structure design with $150^0$ phase advance per cell. The idea of going to a higher phase advance is to reduce the group velocity while keeping a large structure aperture[5]; this may reduce the damage due to rf breakdown [6] without increasing the wakefields. The structure parameters are shown in Table 1.

Table 1. $150^0$ X-band structure parameters

| | |
|---|---|
| F = 11.424 GHz | $T_f$ = 127 ns |
| $\phi$ = $150^0$/cell | $\tau$ = 0.544 |
| $L_{cell}$ = 0.0109343 m | $F_{1,center}$ = 15.21 GHz |
| $L_{struct}$ = 1.4 m | $G_{ave}$ = 6.22 MV/m/(MW)$^{1/2}$ |
| $N_{cell}$ = 128 | $a_{w\perp}$ = 4.74 mm |

For high efficiency, the structure will use rounded cell profiles, which have 12% higher $Q_0$ and shunt impedances than that in the standard disk-loaded waveguide structures. In the locally damped structure, the dipole wakefield generated by the beam is coupled out via four higher-order-mode (HOM) couplers, illustrated in Fig. 1. The coupler wave-guide propagates only the dipole mode frequencies and is cut-off at the fundamental mode frequency. The coupler is narrower in height than the cell length and is positioned off the symmetry plane of the cell in the axial direction so that modes which have TE(M)111-like symmetry can also be damped.

The damping properties of the cell are determined by the coupling iris to the damping waveguide. To determine the damping requirements for the structure, numerical simulations using LIAR[7] were performed to study the

---

[*]Work supported by the DOE, contract DE-AC03-76SF00515.

BBU as a function of the dipole damping. The wakefields were calculated from the *uncoupled* frequencies and kick factors for the structure parameters in Table 1. This simple model must be updated in the future with a wakefield calculated from the coupled mode frequencies and kick factors however it provides a good estimate of the expected performance. The tracking was done for two different beam configurations: 95 bunches having a 2.8 ns bunch spacing and $1.0 \times 10^{10}$e/bunch; 190 bunches having a 1.4 ns bunch spacing and $0.75 \times 10^{10}$e/bunch. In both cases, the train length was 266 ns and the initial oscillation was $1\sigma_y \approx 3\mu m$ at 10 GeV with $\beta_y \approx 5.7$ m in standard NLC linac. In addition, both were tracked using 1 slice per bunch – this eliminates the effect of the initial energy spread and the short-range wakefield which will reduce the BBU. The results are listed in Table 2 and indicate that a $Q_1$ of less than 750 is acceptable for the X-band design. With a $Q_1$ of over 1000, the emittance of the beam starts to deteriorate quickly.

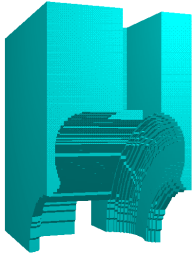

Figure 1. Locally damped structure cell with four damping waveguides; ¼ of the geometry is shown.

Table 2. Results of tracking a $1\sigma_y$ oscillation with (95 bunches of $1.0 \times 10^{10}$e / 190 bunches of $0.75 \times 10^{10}$e).

| $Q_1$ | $\Delta\varepsilon/\varepsilon$ (axis) [%] | $\Delta\varepsilon/\varepsilon$ (centroid) [%] | $W_{sum}$ (rms) [V/pC/m/mm] | $W_{sum}$ (std dev) [V/pC/m/mm] |
|---|---|---|---|---|
| 250 | 47 / 47 | 0.4 / 0.2 | 0.48 / 0.26 | 0.06 / 0.06 |
| 500 | 47 / 47 | 1.0 / 0.5 | 0.52 / 0.38 | 0.08 / 0.09 |
| 750 | 50 / 48 | 2.7 / 1.5 | 0.62 / 0.51 | 0.12 / 0.12 |
| 1000 | 64 / 55 | 13 / 6.9 | 0.94 / 0.77 | 0.28 / 0.24 |

The damping $Q_1$ of the dipole mode was studied using MAFIA [8]. The HOM coupler waveguide dimensions for the model are 12 mm by 3 mm. The coupling iris opens the full height of the HOM waveguide but the width of the iris is adjusted to obtain the required $Q_1$ values while the thickness was kept at 0.6mm. The Q of the dipole mode is very sensitive to the iris opening as is shown in Fig. 2. It is also clear from the plot that very low Q's can be obtained with the local damping scheme.

However, the coupling iris also perturbs the fundamental mode, degrading the fundamental $Q_0$ and the shunt impedance. The degradation is almost inversely proportional to the dipole $Q_1$. The cause of the $Q_0$ reduction is the higher wall-loss in a small region around the iris opening resulted from the current concentration due to the iris perturbation. This wall-loss power may cause excessive local pulsed heating, which must be minimized to avoid damage from the thermal stresses.

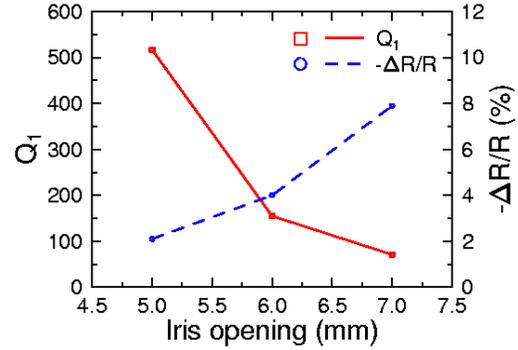

Figure 2. Dipole $Q_1$ and shunt impedance reduction at different iris openings.

## 3 PULSED HEATING

The $Q_0$ reduction of the fundamental mode is due to the perturbation of the HOM coupling iris to the fundamental mode current. The power-loss distribution of the medium damped structure was calculated using Omega3P[9] running on NERSC T3E and SP2 supercomputers. A snap-shot of the wall-loss is shown in Fig. 3. The red spot in a small region around the HOM coupler indicates higher power density which can cause an excessive temperature rise during the rf pulse. A temperature rise of $120^0$C has been shown [10] to be detrimental to the copper surface. Thus, it is important to design the structure to maintain the heating at a safe level. Knowing the power density from the Omega3P simulation, the temperature rise can be calculated using the formula given in [10,11,12]

$$\Delta T = R_s \left| H_{wall} \right|^2 \sqrt{\frac{t_{pulse}}{\pi \rho c k}}$$

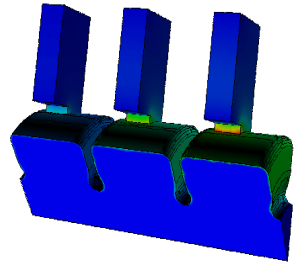

Figure 3. Wall loss distribution

where $R_s = 1/\sigma\delta$ is the surface resistance (0.0279Ω at 11.424 GHz for copper), $H_{wall}$ is the magnetic field on the wall (proportional to the wall current), $T_{pulse}$ is the rf pulse length, $\rho$ is the density of copper (8900 kg/m$^3$), $c$ is the specific heat of copper (385.39 J/kg$^0$C), and $k$ is the thermal conductivity (380 W/$^0$C-m). The temperature rise scaled to a gradient of 70 MV/m is shown in Table 3. The temperature rise depends on both the dipole $Q_1$ and the iris geometry. Smaller iris openings have the least perturbation on the fundamental mode. Rounding the iris corners can further reduce the ΔT.

Table 3. Temperature rise versus iris geometry

| Iris geometry | Dipole $Q_1$ | ΔT |
|---|---|---|
| 3mmx6mm square | 150 | 127 |
| 3mmx6mm rounded | 430 | 80 |
| 3mmx5mm square | 255 | 71 |

The design strategy is to damp the dipole mode only as much as needed to prevent BBU thereby minimizing the loss in $Q_0$ and shunt impedance. This approach leads to a medium damping with a $Q_1$ of a few hundred. We chose a $Q_1$ of 500 for the current medium damped structure design. With such a scheme, it is expected that the rf heating temperature rise will be acceptable.

## 4 HOM LOAD FOR MDS

The loads for the medium damped structure should be simple and compact in geometry. The ideal load would be an adiabatically tapered load which has the advantages of a broad bandwidth and an insensitivity to the material variation. However, the adiabatically tapered load is long; at the X-band dipole frequency, the load would be roughly 50 mm. The disk diameter with such loads would be at least 155 mm which becomes cumbersome as shown in Fig. 4.

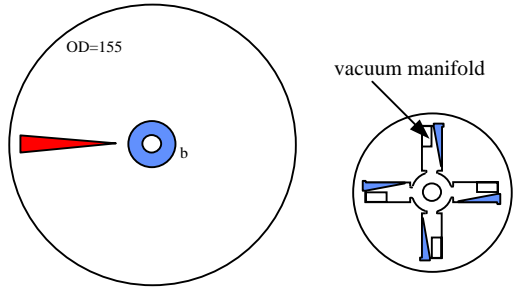

Figure 4. left) adiabatically tapered load, cell OD>155mm; right) compact load using quasi-adiabatic quasi-resonant approach, fits in cell OD of 85 mm.

We took an approach of using quasi-adiabatic quasi-resonant tapers for the loads. Using either MAGO (ε= 9.2(1-j0.13)) or AlN+SiC (ε=25(1-j0.35)) lossy materials, the load can be designed to be shorter than 20 mm. With these loads, the cell can have diameter of 85 mm. For example, the cell geometry with the MAGO load is shown in Fig. 4.

As anticipated, the compact load design has a finite bandwidth which is shown in Fig. 5. Full cell simulations, which include the cell, the HOM waveguide and the loads, suggested that a reflection better than 0.1 is adequate for the medium damped structure. The bandwidth of the load within this reflection limit is shown to be about 1.0 GHz. To cover the whole bandwidth of the dipole modes in the structure, a few different load designs are needed along structure. Simulations show the load is also effective at damping the higher band frequencies (Fig. 5).

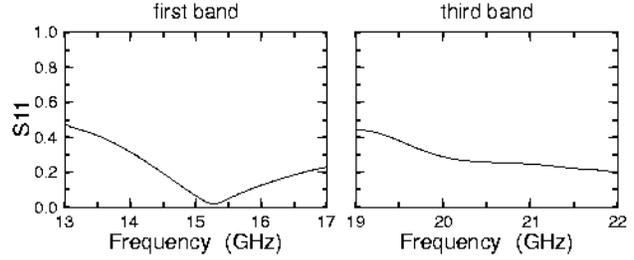

Figure 5. Load reflection at first and third dipole band frequencies.

## 5 VACUUM AND STRUCTURE BPM

Vacuum manifolds can be easily incorporated to the design. A 6mmx10mm manifold, shown in Fig. 4, cutting through the damping waveguide provides additional longitudinal vacuum conductance with little effect on the load design.

Unlike the RDDS type structures, the structure BPM for the locally damped structure is not readily available. The present thought is to utilize probes in the damping waveguides to monitor the dipole signals. A few such probes are needed to determine the alignment of the structure. Another option is to couple the vacuum manifold weakly to the rf in the damping waveguide and pick up the signals in the manifold for beam position analysis. Further studies are needed on these issues.

## ACKNOWLEGEMENT

The authors would like to thank G. Bowden, R. Ruth, P. Wilson and R. Miller for helpful discussions on pulse heating of accelerator structures.